\documentclass[twocolumn]{aastex631}

\shorttitle{Precise physical properties of OGLE-LMC-CEP-1347}
\shortauthors{Espinoza-Arancibia et al.}

\begin{document}

\title{A novel q-PED method: precise physical properties of a merger-origin binary Cepheid OGLE-LMC-CEP-1347}

\author[0000-0003-0264-1234]{Felipe Espinoza-Arancibia}
\email{fespinoza@camk.edu.pl}
\affiliation{Centrum Astronomiczne im. Mikołaja Kopernika, PAN, Bartycka 18, 00-716 Warsaw, Poland}

\author[0000-0003-3861-8124]{Bogumił Pilecki}
\affiliation{Centrum Astronomiczne im. Mikołaja Kopernika, PAN, Bartycka 18, 00-716 Warsaw, Poland}

\begin{abstract}

Recently, a double-lined binary (SB2) classical Cepheid, OGLE-LMC-CEP-1347, was discovered, with the orbital period (P$_{\rm orb} = 59$ days) five times shorter than of any binary Cepheid known before. The expected mass of the Cepheid was below $3.5$ M$_\sun$, which, if confirmed, would also probe the uncharted territory. The system configuration also pointed to the Cepheid as a merger. We present a novel method for determining precise physical parameters of binary Cepheids using both theory and observations. This q-PED method combines the measured mass ratio (q), pulsation (P), and evolutionary (E) models, and the known distance (D) supplemented with multi-band photometry. Applying it, we determined the mass of the Cepheid of $3.41 \pm 0.08$ M$_\sun$, its radius of $13.65 \pm 0.27$ R$_\sun$, the companion mass of $1.89 \pm 0.04$ M$_\sun$ and radius of $12.51 \pm 0.62$ R$_\sun$. With the current configuration, the apparent evolutionary age difference of almost 1 Gyr between the components strongly favors the Cepheid merger origin scenario. If so, the actual age of the Cepheid would be 1.09 Gyr, on the edge of Population II stars, indicating a significant fraction of Cepheids may be much older than typically assumed. We also applied our method to an eclipsing binary Cepheid OGLE-LMC-CEP-1812 with accurately determined physical parameters, obtaining a close agreement, which confirmed our method's reliability.

\end{abstract}


\section{Introduction} \label{sec:intro}
Classical Cepheids (hereafter Cepheids) are pulsating giants and supergiants that occupy a limited area on the Hertzsprung-Russel diagram (HRD) known as the instability strip (IS). These stars exhibit a tight relation between their pulsation period and luminosity, making them essential for calculating extragalactic distances. Additionally, Cepheids are an indispensable tool for testing predictions of stellar evolution and pulsation theory. Typically, Cepheids first enter the IS during the H-shell burning phase after they have evolved beyond the main sequence phase. The star crosses the IS again during the He-core burning phase, called the blue loop. This evolutionary phase is strongly influenced by metallicity and the chosen input physics, such as convective overshooting, nuclear reactions, and rotation \citep[see, e.g.,][]{2004Xua,2004Xub, Zhao2023, Smolec2023, Ziolkowska2024}. The time scale of the first crossing is about 100 times shorter than subsequent crossings. As a result, most observed Cepheids are generally believed to be located within blue loops.

Cepheids span a typical mass range between 3 and 13 M$_\sun$ \citep[see, e.g.,][]{Bono2000, Anderson2016}. For low-mass stars ($\lesssim$ 3.5 M$_\sun$), the extension of blue loops predicted by evolutionary models is too short to reach the IS and develop the right conditions for the star to pulsate as short-period Cepheids \citep{Anderson2016, DeSomma2021, Espinoza2022, Espinoza2024}. Currently, the existence of short-period Cepheids can only be explained if they are found to be crossing the IS for the first time \citep{Ripepi2022, Espinoza2024}. To verify this scenario and improve our understanding of the processes governing blue loops, measurements of short-period Cepheid dynamical masses are necessary. However, all seven dynamical masses measured so far fall within a narrow mass range near 4 M$_\sun$ \citep{Pilecki2018, Gallenne2018}.

Recently, an interesting spectroscopic binary system with a double-mode Cepheid (OGLE-LMC-CEP-1347, henceforward CEP-1347) was discovered \citep{Pilecki2022}. This system is remarkable for having the shortest orbital period (P$_{\rm orb} = 59$ days) among all known binary Cepheids, and the Cepheid itself has the shortest pulsation periods ever observed in such systems. Its first-overtone (1O) period is P$_{\rm 1O} = 0.690$ days, and for the second-overtone (2O) it is P$_{\rm 2O} = 0.556$ days. These short periods and fast period change suggest that CEP-1347 is on the first crossing of the IS \citep{Pilecki2022}. Additionally, the evolutionary analysis by \citeauthor{Pilecki2022} showed that the Cepheid is probably a product of a merger of two less massive stars. The unique characteristics of this system make it an essential candidate for mass and evolutionary state determination. Moreover, strong constraints on the properties of this and similar systems can shed light on the evolutionary origin of short-period Cepheids. 

In this letter, we introduce a novel q-PED method that combines theory and observations, using the measured mass-ratio ($q$), pulsation (P), and evolutionary (E) models, and the known distance (D), in addition to multi-band photometry, to constrain the physical parameters of the Cepheid and the companion of CEP-1347. In Sect.~\ref{sec:modeling}, we described the different steps of the q-PED method used in this work to model the binary system CEP-1347. In Sect.~\ref{sec:discussion}, we discuss the evolutionary status of CEP-1347, and we test our method for a well-measured eclipsing binary Cepheid, OGLE-LMC-CEP-1812, which exhibits an evolutionary similar binary configuration.

\section{The \MakeLowercase{q}-PED method} \label{sec:modeling}
Our method consists of three main parts. First, we calculate an initial grid of evolutionary tracks for the Cepheid and its companion using the measured mass ratio of the system. Second, we compute linear pulsational models for every point along the Cepheid's evolutionary track within the IS, and we select those pulsating with the same P$_{\rm 1O}$ and P$_{\rm 2O}$ values as CEP-1347. Finally, we use the distance of the LMC as a constraint to identify valid system configurations of the Cepheid pulsational models and positions of the companion on corresponding evolutionary tracks.

\begin{figure}[!ht]
    \centering
    \includegraphics[width=\hsize]{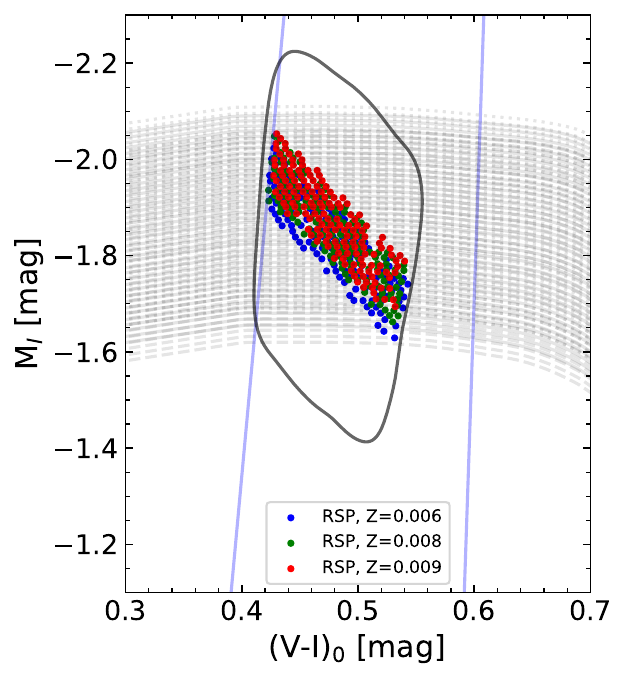}
    \caption{Color-magnitude diagram showing RSP models that exhibit 1O- and 2O-mode pulsations with periods within five percent from measured $P_{\rm 1O}=0.690$ days and $P_{\rm 2O}=0.556$ days, considering three values of metallicity. These models were calculated along the evolutionary tracks with metallicity $Z=0.006$ (dashed lines), $0.008$ (solid lines), and $0.009$ (dotted lines). Blue solid lines are the empirical IS edges for 1O Cepheids \citep{Espinoza2024}. The black solid line shows the contours of RSP models that exhibit 10- and 20-mode pulsations within 20\% from the measured periods.}
    \label{fig:HRD_2}
\end{figure}

As we do not know the masses of the components, to perform the first part of our method, we first had to find the possible mass range for the Cepheid. This was done in a few steps. The first one was to estimate the luminosity range for CEP-1347, using calibrated I-band period-luminosity relation for fundamental (F) mode Cepheids from \cite{Breuval2022}. To use this relation, the 1O-mode period of CEP-1347 was fundamentalized \citep{Pilecki2024L} using the fundamentalization equation given in \cite{Pilecki2021}, yielding $P_{\rm F}=0.966$ days. For this period, the estimated absolute I-band magnitude is $M_I=-1.74\pm0.01$ mag. This value was compared with sparsely calculated preliminary evolutionary tracks to obtain a rough estimate of the Cepheid mass, $M_{P-L} \sim$ 3.3-3.4 M$_{\odot}$. Subsequently, we computed a dense grid of evolutionary tracks with a safe margin around $M_{P-L}$, covering masses from $3$ M$_{\odot}$ to $4$ M$_{\odot}$ in steps of $0.01$ M$_{\odot}$. We used the version r22.11.1 of the code Modules for Experiments in Stellar Astrophysics \citep[MESA;][]{Paxton2011,Paxton2019,Jermyn2023} adopting three representative metallicity values  $Z=0.006$, $0.008$, and $0.009$ for LMC Cepheids \citep{Romaniello2022}. Additionally, using the MESA functionality Radial Stellar Pulsations \citep[RSP;][]{Paxton2019}, we calculated pulsation models on each point of the tracks inside the empirical IS borders for 1O mode pulsations from \cite{Espinoza2024}. We then constrained the mass range of the models by selecting the evolutionary tracks that presented RSP models with 1O and 2O-mode periods equal to the observed ones (P$_{\rm 1O}=0.690$ d and P$_{\rm 2O}=0.556$ d) within 5\% margin.
This way, the Cepheid tracks were limited to $3.15$ to $3.65$ M$_\sun$. We then used the system's measured mass ratio $q=0.553$ \citep{Pilecki2022} to compute a grid of tracks for the companion star covering a mass range of $1.74$ to $2.01$ M$_\sun$. 
Subsequently, we required positive 1O and 2O-mode growth rates for the Cepheid to ensure both modes were excited. Effectively, from the blue side, the models are restricted by the blue edge of the 1O-mode IS, and from the red side, the IS red edge for the 2O-mode. The possible Cepheid positions on the HRD are shown in Figure~\ref{fig:HRD_2}.

To test how the unknown uncertainties present in the applied theoretical models may affect the results, we performed the above-described procedure with a 20\%-tolerance of RSP model periods -- the outcome of this test is included in Sec.~\ref{sec:discussion}.

Our evolutionary models consider the solar mixture provided by \cite{Grevesse1998}. We used a mixing length parameter of $\alpha_{\rm mlt} = 1.939$. For the convective boundaries, we used the predictive mixing scheme described in \cite{Paxton2018}, in addition to exponential core and envelope overshooting with parameters $f_{c,ov} = 0.015$ and $f_{en,ov} = 0.024$, respectively. The evolutionary tracks account for mass loss during the red giant branch (RGB) phase, using the \cite{Reimers1975} prescription, with a scaling factor $\eta_R = 0.3$. A comprehensive study of the impact of the choice of different input physics on evolutionary tracks can be found in \citet{Ziolkowska2024}. Their conclusions indicate that for models with convective overshooting, the mass- and metallicity-averaged uncertainties in $(\log T_{\rm eff},\log L$) due to the use of different input physics, at the terminal age main sequence and the end of the RGB are $(0.2\%, 1\%)$, and $(0.3\%, 1.5\%)$, respectively.

On the other hand, we limited our RSP pulsation models to linear calculations. The non-linear nature of double-mode pulsations is not yet implemented in RSP and its detailed modeling remains an open problem \citep{Smolec2008a}. The RSP model depends on free parameters in equations describing time-dependent convection from \cite{Kuhfuss1986}, implemented by \cite{Smolec2008b}. These parameters need calibration for different types of pulsating stars. Currently, there is no calibration for classical Cepheids \citep{Kovacs2023}, and the approximate parameter sets from \citet{Paxton2019} are used in the literature \citep{Das2020,Kurbah2023,Hocde2024,Deka2024}. We adopted the convective free parameters of set D. In a few cases where the RSP calculations did not converge, we changed the convective parameters to the ones of set C, which is the most similar to set D. The MESA inlists used in this work can be downloaded from Zenodo\dataset[10.5281/zenodo.13971570]{https://doi.org/10.5281/zenodo.13971570}.

\begin{figure*}[!ht]
    \centering
    \includegraphics[width=\hsize]{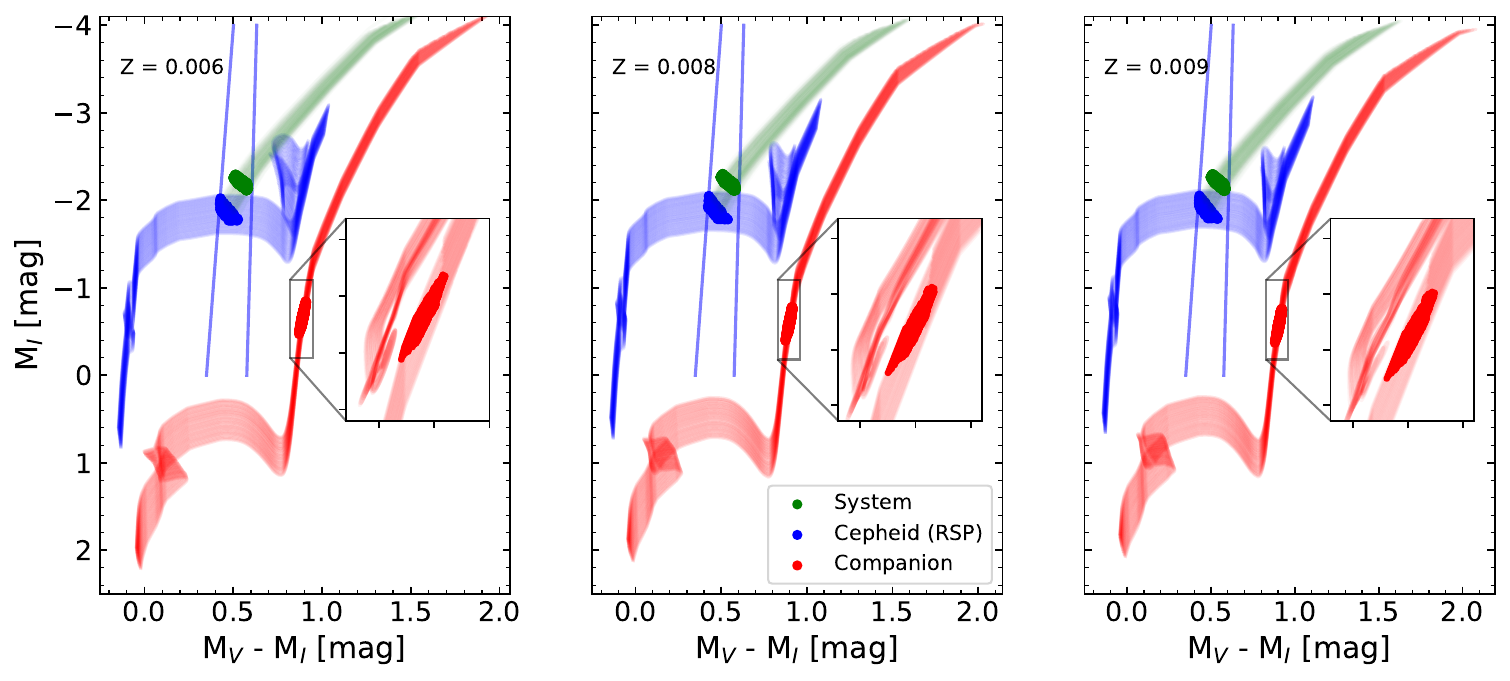}
    \caption{Color-magnitude diagrams with the evolutionary tracks of the Cepheid (blue), the companion (red), and the entire CEP-1347 system (green). Valid positions for the Cepheid, companion, and the entire system are shown as blue, green, and red points, respectively.}
    \label{fig:HRD_3}
\end{figure*}

We computed the possible location in the HRD of the CEP-1347 system by adding up (logarithmically) the Cepheid magnitudes of the RSP models shown in Figure~\ref{fig:HRD_2} with each point of the companion's evolutionary track. The resulting tracks, to which we will refer as system tracks, are shown as green lines in Figure~\ref{fig:HRD_3}. The final position of the system in the HRD was determined by constraining the system tracks to the known distance to the LMC \citep{Pietrzynski2019}. To make this constraint more robust, we used the multi-band method similar to that applied to Cepheids in \citet{Gieren2005}. In general, this method consists of fitting the relationship
\begin{equation}
\label{eq:1}
(m-M)_0 = (m-M)_\lambda - E_{B-V}R_\lambda,
\end{equation}
with the known values of the total to selective absorption $R_\lambda$ in the VIJHK-bands \citep[3.057, 1.777, 0.812, 0.508, and 0.349 respectively; from][]{Breuval2022}, and the reddened distance moduli $(m-M)_\lambda$. The resulting intercept and slope of the linear fit provide an estimate of the true distance modulus $(m-M)_0$, hence the system's distance, and the reddening $E_{B-V}$.

\begin{figure}[!ht]
    \centering
    \includegraphics[width=\hsize]{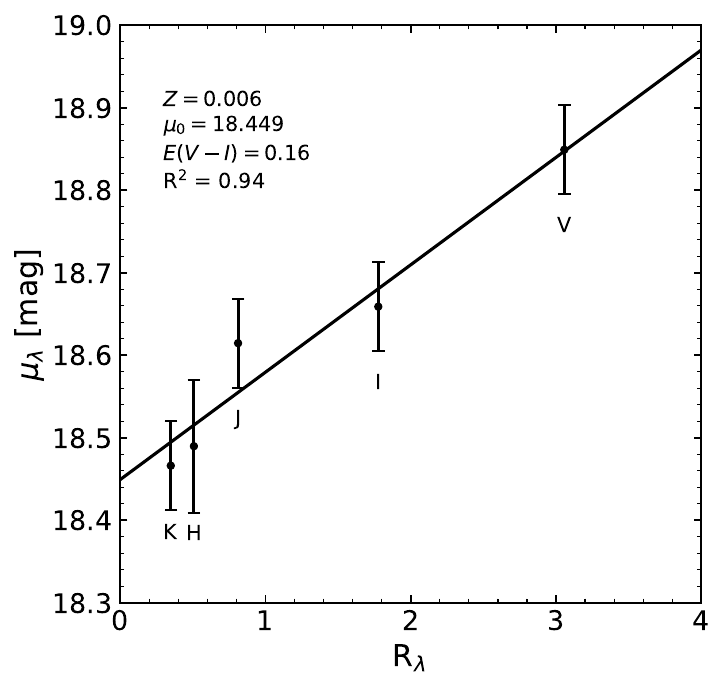}
    \caption{Multi-band fit with the highest R-squared of the reddened distance moduli in the VIJHK bands as a function of the selective absorption $R_\lambda$. }
    \label{fig:multiband}
\end{figure}

To compute the reddened distance moduli of CEP-1347, we needed apparent and absolute magnitudes of the system. For the apparent ones, we used values and uncertainties, in VI-bands from OGLE IV \citep{Soszynski2015}, in JK-bands from \citet{Ripepi2022} transformed to the 2MASS photometric system using the equations provided by \citet{Breuval2021}, and in H-band from the 2MASS 6X Point Source Working Database \citep{Cutri2012} by a search within 1\arcsec~using the VizieR catalog access tool. Regarding the absolute magnitudes, in the original method, values from fiducial P-L relations for Cepheids are taken, but in our case, we need the total magnitude for the system. To obtain it, we used the information from the evolutionary tracks of the system. These tracks were transformed to absolute VIJHK-bands magnitudes using two sets of bolometric corrections, one included in MESA \citep{Paxton2018}, based on \citet{Lejeune1998}, and another from \citet{Choi2016} for JHK-bands. We compared the absolute magnitudes in the V- and I-band at the first IS crossing, calculated with these two sets, and found average differences of $0.05$ mag. We considered these differences as a systematic uncertainty to the absolute magnitudes.
We computed the reddened distance modulus for every point of the system tracks on the different bands using the apparent and absolute magnitudes and their propagated uncertainties.
We iteratively applied the multi-band method to all reddened distance moduli, limiting our results to $(m-M)_0=18.487\pm 0.04$ \citep{Pietrzynski2019}, positive reddening, and an R-squared parameter (which provides information about the goodness of fit of the linear model) higher than $0.8$. The points on the system tracks (and the corresponding positions of the Cepheid and companion star) whose results were within the defined constraints were considered valid configurations of the CEP-1347 system.
However, for the companion, we excluded the models after the RGB evolution because they are incompatible with the current configuration (orbital period longer than 200 days would be needed; mode detailed justification is given in Section~\ref{sec:discussion}).
The final position in the HRD of the components of CEP-1347 is shown in Figure~\ref{fig:HRD_3}. As an example, the fit of the multi-band distance moduli with the highest R-squared obtained for the models with $Z=0.006$, is shown in Figure~\ref{fig:multiband}. The mean physical parameters of the Cepheid and its companion and their standard deviation considering RSP models within 5\% and 20\% are summarized in Table~\ref{tab:1}. 

\begin{deluxetable*}{cccccc}
\tablecaption{Physical parameters of CEP-1347.\label{tab:1}}
%\tablewidth{0pt}
\tablehead{
\colhead{Parameter} & \colhead{Cepheid ($5 \%$)} & \colhead{Cepheid ($20 \%$)} & \colhead{Companion ($5 \%$)}& \colhead{Companion ($20 \%$)} & \colhead{Unit}}
\startdata
Mass & $3.41 \pm 0.08$ & $3.43 \pm 0.13$ & $1.89 \pm 0.04$ & $1.92 \pm 0.07$ & M$_\sun$ \\
Radius & $13.65 \pm 0.27$ & $13.97 \pm 0.89$ & $12.51 \pm 0.62$ & $11.68 \pm 1.67$ & R$_\sun$ \\
log g & $2.705 \pm 0.013$ & $2.689 \pm 0.044$ & $2.520 \pm 0.047$ & $2.594 \pm 0.138$ & cgs \\
Temperature & $6490 \pm 96$ & $6452 \pm 122$ & $4910 \pm 63$ & $4967 \pm 120$ &K\\
log L & $2.47 \pm 0.03$ & $2.48 \pm 0.05$ & $1.91 \pm 0.03$ & $1.86 \pm 0.09$ & L$_\sun$\\
Age & $0.23 \pm 0.01$ & $0.23 \pm 0.02$ & $1.09 \pm 0.07$ & $1.07 \pm 0.1$ & Gyr \\
$E_{\rm B-V}$ &\multicolumn{4}{c}{$0.08\pm0.02$} & mag\\
Distance$^1$  &\multicolumn{4}{c}{$49.8\pm0.5$} & kpc\\
\enddata
\tablecomments{$^1$ Used as a constraint.}
\end{deluxetable*}

\section{Discussion} \label{sec:discussion}

The precise results of combining RSP pulsation models with the multi-band method are in agreement with the hypothesis of the merger origin of the Cepheid component of the CEP-1347 system proposed by \cite{Pilecki2022} since the age difference of the two components is close to 1 Gyr. The Cepheid mass obtained in our analysis is lower than any dynamically measured Cepheid mass to date. The companion is a low-mass star, fainter and redder than the Cepheid, although, interestingly, of similar size. It is essential to mention that the obtained age of the Cepheid is based on its current mass and single-star evolution, while the age of the binary system corresponds to the age of the companion star. Most likely, the system was a triple before, and the Cepheid is a merger of inner binary components with masses close to half the current mass of the Cepheid. Following \citet{Glebbeek2008, Glebbeek2013}, for central collisions between low-mass stars, a fraction of mass of about $10 \%$ is ejected by the collision. This implies that the initial system would have a mass of around $3.8$ M$_\sun$, and each component would be a $\sim 1.9$ M$_\sun$ star. These components would evolve at a rate similar to that of the current companion to the Cepheid. Because we model the merger product using the single-star evolution, in this study, we assume that the merger happened during the main sequence evolution, and the resulting internal structure is equivalent to a non-interacting star. Our results may not be fully representative if other merger scenarios had to be considered. In the future, it would be highly valuable to study their effect on the current structure and composition of the Cepheid and verify them once the available data permits. 

If the Cepheid is indeed of merger origin, its actual age is about 1.1 Gyr. Considering the observational conclusion of \citet{Pilecki2024} about the relatively high rate of possible mergers among binary Cepheids \citep[also indicated by recent theoretical simulations of][]{Dinnbier2024}, this implies that a significant fraction of Cepheids is older than they appear and may even belong to Population II.

Technically, other mechanisms can lead to a binary having components of significantly unequal ages, either apparent or actual. One of them is mass transfer, but it is improbable that it could lead to the system's current configuration. As shown in simulations of \citet{Neilson2015-2}, interactions of components during the RGB evolution cause Cepheids in binaries to have orbital periods longer than 200 days. And given the current configuration of CEP-1347, any binary interaction could not happen without the initially more massive component evolving on the RGB. Its orbital period is, however, only about 59 days. There is also another reason to exclude the mass transfer. Typically, such events rejuvenate the less massive star, which is already less evolutionary advanced, and produce a pair of a main sequence star (gainer) and an evolutionary advanced envelope-stripped star (donor). For mass to be transferred to the current Cepheid, the companion star should have been more massive and evolved more rapidly to fill its Roche lobe.  Given the current mass ratio, a considerable amount of mass should have been transferred, at least $0.8$ M$_\sun$. This implies that the current Cepheid would have gained more than $30\%$ of its original mass, mainly in hydrogen. Its further evolution on the main sequence would be much slower than that of the evolved donor, having difficulty catching up with the current companion. A good example of a system after the mass transfer between the components is OGLE LMC-T2CEP-0211 \citep{Pilecki2018b}.

Another mechanism that can be considered is a stellar capture scenario \citep{Heggie2003}. However, using a population synthesis code ({\rm StarTrack}), \citet{Ivanova2005} showed that the total number of tidal-capture binaries formed during a globular cluster lifetime is less than 1 percent of the final number of binaries in its core. This is already significantly less than the observed fraction of binary Cepheids with anomalous mass ratios of at least 20\% \citep{Pilecki2018, Pilecki2024}. Nevertheless, we would like to note that CEP-1347 is not part of any known stellar cluster, and even if this were the case, the captured star should be of similar age. This means that field stars should have to be involved, which decreases the chance of the capture scenario to a negligible value. Moreover, the system's orbit is almost circular, suggesting a longer binary evolution that circularized the orbit, while stellar capture would preferentially produce highly eccentric systems \citep[see, e.g.,][]{Hamilton2024}.
Therefore, until disproved, we treat the merger origin of the Cepheid as the most plausible scenario consistent with all available observational and theoretical results. Nevertheless, as our method includes several observational constraints, the results should be fairly independent of the scenario and even more correct for the stellar capture (single-star evolution is used in the method). Even in the case of the mass transfer, only the structure and chemical composition of the companion would be more significantly affected, with little influence on the Cepheid.
With no constraint on the companion's evolutionary phase, our method also leads to valid models at the core He-burning phase. However, the short orbital period (P$_{\rm orb} = 59$ days) and low component separation, $a\approx 107$ R$_{\sun}$ \citep{Pilecki2022}, make them very unlikely since the companion star would increase its size considerably evolving towards the tip of the RGB. The mass transfer from the companion to the Cepheid would not stop until the orbit grows to more than twice the observed one, with the period reaching at least 200 days. More details regarding the exclusion of the mass transfer have already been given above. For this reason, we excluded the models with the companion at the core He-burning phase from our analysis.
The compactness of the binary suggests that the Cepheid may interact with its companion in the future \citep{Pilecki2022}. Our evolutionary tracks show that in around $1.4$ Myr, the system's primary component will evolve to the tip of the RGB (no longer a Cepheid), and its size will be about $3$ times larger. During the same time, the companion will remain of similar size, but eventually, it will also grow to about 90 $R_\odot$ while the primary will be on the blue loop. This suggests that binary interactions between both components will determine the future evolution of the CEP-1347 system, with an additional merger being a likely scenario.

\begin{deluxetable*}{cccccc}
\tablecaption{Comparison of physical parameters of OGLE-LMC-CEP-1812.\label{tab:2}}
\tablehead{& \multicolumn{2}{c}{Cepheid} & \multicolumn{2}{c}{Companion} &\\
Parameter&This work&\citet{Pilecki2018}&This work&\citet{Pilecki2018}&Unit}
\startdata
Mass & $3.75 \pm 0.08$ & $3.76 \pm 0.03$ & $2.63 \pm 0.05$ & $2.62 \pm 0.02$  & M$_\sun$ \\
Radius & $17.30 \pm 0.35$ & $17.85 \pm 0.13$ & $12.13 \pm 0.70$ & $11.83 \pm 0.08$ & R$_\sun$ \\
log g & $2.542 \pm 0.014$ & $2.509 \pm 0.007$ & $2.692 \pm 0.055$ & $2.709 \pm 0.007$ & cgs \\
Temperature & $6233 \pm 94$ & $6120 \pm 150$ &  $5257 \pm 101$ & $5170 \pm 120$ &K\\
log L & $2.60 \pm 0.03$ & $2.61 \pm 0.04$ & $2.00 \pm 0.05$ & $1.95 \pm 0.04$ & L$_\sun$\\
Age & $0.182 \pm 0.007$ & $0.190$ & $0.487 \pm 0.046$ & $0.369$ & Gyr \\
\enddata
\end{deluxetable*}

A well-known problem in Cepheid modeling is that pulsation modeling or dynamical determinations of the Cepheid mass are smaller than those determined from stellar evolution modeling \citep[see, e.g.,][]{Keller2008}. Although the situation has improved in recent years, and it was shown that the discrepancy could be removed for individual objects \citep{Pietrzynski2010,Cassisi2011,PradaMoroni2012}, there is no consensus regarding the mechanism responsible for the difference. Therefore, the solution has not yet been universally introduced to evolutionary codes \citep[e.g.,][]{Gallenne2018}. Since our method relies on evolutionary models, to evaluate biases in our results, we analyzed the eclipsing binary Cepheid OGLE-LMC-CEP-1812 (hereafter CEP-1812) in the same way. This system has precise orbital and physical parameters \citep{Pilecki2018} and, similarly to CEP-1347, it has been suggested that the Cepheid is a merger of two lower-mass components \citep{Neilson2015}.
Moreover, the components of the system have the same evolutionary state as CEP-1347, with an F-mode Cepheid at the first crossing of the IS and a less massive companion on the RGB. This hints that both systems were part of the same class of triple systems. We applied our method using VI-bands magnitudes from OGLE III \citep{Soszynski2008}, JHK-bands magnitudes from \citet{Inno2016}, and MESA evolutionary tracks in a mass range of $3.47$ to $4.0$ M$_\sun$. The results of this analysis are shown in Table~\ref{tab:2}. Considering the abovementioned problems, the agreement with the results obtained from the orbital modeling by \cite{Pilecki2018} is excellent, especially regarding the mass. The most significant difference is for the radius, with our value being $3\%$ (1.5$\sigma$) smaller. Although more test objects would be necessary to estimate systematic errors of our method reliably, these results suggest they are small and generally should not exceed 4\%.

\cite{Pilecki2018} established a period-mass-radius (PMR) relation based on accurate physical parameters measured for 6 Cepheids in eclipsing binary systems. Applying this relation to the mass and radius determined for CEP-1347, we obtained a fundamental-mode period $P_{\rm F,PMR} = 0.90 \pm 0.10$, that is 4\% shorter but consistent within errors with the fundamentalized period of CEP-1347, $P_{\rm F} = 0.934 \pm 0.003$ days. Note that this time, for fundamentalization, we used eq.~1b from \cite{Pilecki2024L}\footnote{Using the relation standard deviation, $\sigma=0.004$, provided therein for the calculation of uncertainty.}, which is more appropriate when comparing physical properties. This result is consistent with our selection of RSP pulsation periods of 5\% around the observed 1O and 20-mode periods using period values only, and of 20\% when uncertainties are considered. On the other hand, applying this relation to the mass determined for CEP-1347 and its fundamentalized period, we obtained a radius of 13.95 R$_{\odot}$, which is $2.2\%$ larger than the one obtained using our q-PED method. This discrepancy aligns with the results for CEP-1812, indicating a possible underestimation of radii in our codes. The difference may be associated with the models and the binary origin of the Cepheids in these systems. An independent investigation on this subject will be part of another work.

As an additional test, we used the theoretical PMR relation based on the nonlinear, convective pulsation models from \citet{Somma2022}. We selected their relation for 1O-mode Cepheids, considering their case B of mass-luminosity relation, a convection efficiency of $\alpha_{\rm ml} = 1.7$, and metallicity of $Z=0.008$, given that their IS computed with this set of parameters showed good agreement with the LMC empirical IS \citep{Espinoza2024}. Similar to the previous test, we applied this relation to the mass and radius determined for CEP-1347, yielding $P_{\rm PMR,1O} = 0.676 \pm 0.043$ days. This period is consistent within $0.3\sigma$ with the observed 1O period of CEP-1347, differing by only two percent. This test strengthens our selection of RSP pulsation periods and shows that independent results using nonlinear pulsation models are consistent with the results of our method.

As mentioned in Sec.~\ref{sec:modeling}, we checked the effect on the results of selecting RSP pulsation models with periods in a range increased to 20\% from the measured ones (see Table~\ref{tab:1}). There was no significant change in the mean values, while the uncertainties increased slightly. The only exception was the radius (and dependent on it, $\log g$), which increased by 2\%, with its uncertainty tripled. This change in the radius value and uncertainty makes up for the previously encountered differences. Still, a larger sample of test objects will be needed to check if the effect is statistical (depends on the target) or systematic (depends on the model used).

An essential feature of the CEP-1812 and CEP-1347 systems is that the Cepheids are at the first crossing of the IS. Therefore, we avoid all the complexity of modeling blue loops, which allows us to obtain more accurate results for the system. However, only two such binary systems are known to date, making a statistical analysis of systematic errors impossible. Nevertheless, the excellent agreement of the results of this work with those obtained by \citet{Pilecki2018} shows that this method can be used confidently for Cepheids in double-lined binary systems for which the mass ratio can be measured. Apart from the measured q, the empirical part of the method is strengthened by good-quality photometry, particularly in the NIR bands. These observational constraints form a vital part of the analysis, allowing, for example, an accurate estimate of the luminosity of the system and its components, as seen in Table~\ref{tab:2}.

In summary, the q-PED method proved to be a valuable tool for determining the physical parameters of binary systems and testing evolutionary models. In particular, very accurate masses and luminosities may be used to determine the still poorly defined mass-luminosity relation for Cepheid variables, which is a crucial base for their theoretical understanding. In future studies, we will use this method to study a large sample of more than 20 systems with Cepheids (mainly those evolving through blue loops), which will allow us to enhance the calibration of the method further.

\begin{acknowledgments}
We thank the anonymous referee for the constructive comments and suggestions. We acknowledge support from the Polish National Science Center grant SONATA BIS 2020/38/E/ST9/00486. F.E. thanks Oliwia Zi\'{o}\l{}kowska for her helpful insights in calculating the companion's evolutionary tracks. This research has used the VizieR catalog access tool, CDS, Strasbourg, France \citep{10.26093/cds/vizier}.
\end{acknowledgments}

\bibliography{main}{}
\bibliographystyle{aasjournal}

\end{document}